\documentclass[12pt,amssymb]{article} 

\newcommand{\be}{\begin{equation}}
\newcommand{\ee}{\end{equation}}
\newcommand{\bea}{\begin{eqnarray}}
\newcommand{\eea}{\end{eqnarray}}

\begin{document}

\begin{titlepage}
\begin{center}
\vskip .2in \hfill \vbox{
    \halign{#\hfil         \cr
           hep-th/0111007 \cr
           UCSD-PTH-01-19\cr
           November 2001    \cr
           }  
      }   
\vskip 1.5cm {\Large \bf The Secret Gauging of Flavor Symmetries
in
Noncommutative QFT} \\
\vskip .1in
\vskip .3in {\bf Ken Intriligator,} \footnote{email
address:keni@ucsd.edu} and {\bf  Jason Kumar}
\footnote{e-mail address:j1kumar@ucsd.edu}\\
\vskip .25in {\em  Department of Physics,
University of California, San Diego\\
La Jolla, CA  92093-0354 USA \\} \vskip .1in \vskip 1cm
\end{center}
\begin{abstract}
We show that flavor 't Hooft anomalies {\it automatically} vanish
in noncommutative field theories which are obtained from string
theory in the decoupling limit.  We claim that this is because the
flavor symmetries are secretly local, because of coupling to
closed string bulk modes.  An example is the SU(4) R-symmetry of
N=4 D=4 NCSYM.  The gauge fields, along with all closed string
bulk modes, are not on-shell external states but do appear as
off-shell intermediate states in non-planar processes; these
closed string modes are thereby holographically encoded in the
NCFT. \vskip 0.5cm
\end{abstract}
\end{titlepage}
\newpage

\section{Introduction}
Non-commutative field theories (NCFT) exhibit fascinating
similarities with string theory.  In the realization of
non-commutative field theories via D-branes in a background $B$
field, one sees that the decoupling limit \cite{Seiberg:1999vs}\
retains some of the stringy behavior, which is usually thrown away
in the decoupling limit for commutative QFTs.  In particular,
diagrams with non-planar momentum exhibit UV suppression, which
can lead to UV/IR mixing \cite{Minwalla:2000px,
VanRaamsdonk:2000rr}.  This effect was interpreted in
\cite{Minwalla:2000px, VanRaamsdonk:2000rr} as non-decoupling of
closed string modes (see also \cite{Rajaraman:2000dw}), and can
also be viewed as the effect of stretched open strings
\cite{Sheikh-Jabbari:1999vm} - \cite{Liu:2000ad}.  A way to think
about this \cite{Minwalla:2000px} is that the closed string
propagator involves $(\alpha ')^2 g^{\mu \nu}$, which one might
expect to vanish in the decoupling limit $\alpha '\rightarrow 0$;
however, holding the open string metric fixed to $G^{\mu \nu}=\eta
^{\mu \nu}$, we actually have in the decoupling limit \be (\alpha
')^2 g^{\mu \nu}\rightarrow -(\Theta ^2)^{\mu \nu}.\ee

We claim, based on eqn. (1), that all closed string modes remain
coupled to the NCFT.  In the decoupling limit, the closed string
modes do not exist as on-shell external states, but they still
remain coupled as intermediate, off-shell contributions to
non-planar loop diagrams.  As a particular case of this, we argue
that worldvolume flavor symmetries are secretly gauged, with the
gauge bosons coming {}from closed strings which remain coupled via
eqn. (1).  As evidence, we discuss anomalies.

The UV suppression of non-planar diagrams in NCFT, which is
related to eqn. (1), ensures that no non-planar diagrams are
anomalous \cite{Martin:2000qf,Intriligator:2001yu}.  Thus mixed
anomalies, which could only arise via non-planar diagrams, always
automatically vanish in NCFT.  The interpretation of
\cite{Intriligator:2001yu} is that these anomalies are cancelled
because the NCFT automatically contains the necessary closed
string axion fields which cancel the anomaly by the Green-Schwarz
mechanism, precisely as in string theory.  This can be seen
explicitly in string theory constructions of chiral NCFTs.  As
also pointed out in \cite{Intriligator:2001yu}, all ABJ type
anomalies, involving both flavor and gauge currents, are also
automatically cancelled by a Green-Schwarz type mechanism.

We will here argue that all flavor 't Hooft anomalies also
automatically vanish in NCFTs which are obtained via string
theory. This is caused by the same nonplanar UV suppression
factors which occur via eqn. (1).  Our interpretation of this, as
well as the vanishing of the ABJ anomalies, is that the flavor
symmetries are secretly gauged, and thus necessarily anomaly free.
String theory constructions, of course, will always lead to a
consistent theory.  The 't Hooft anomaly cancellation, and secret
gauging, is via closed string modes, which remain coupled
essentially due to eqn. (1).

The flavor symmetries of the NCFT living in the woldvolume of a
D-brane are the ``normal bundle'' rotations of the directions
transverse to the brane.  Before taking the decoupling limit,
these are gauge symmetries of the bulk theory, which includes
gravity. The worldvolume flavor current couples to bulk gauge
fields, which can be written in terms of the bulk metric.  For
D-branes in a general curved bulk metric background, we'll have
terms in the worldvolume action like \cite{Douglas:1998zw}
$$\int d^{p+1}x str[g_{ab}(\Phi ^a)D_\alpha \Phi ^a D^\alpha \Phi ^b]+
\hbox{fermion terms}.$$ Here $str$ is the symmetrized trace over
the worldvolume $U(K)$ gauge indices, $D_\alpha$ are the $U(K)$
gauge (and not flavor) covariant derivatives, with $\alpha =0\dots
p$ the worldvolume space-time index, and $a,b,c$ run over the bulk
transverse directions. Expanding this around a classical solution
for $\Phi ^a$ gives a coupling \be \int d ^{p+1}x ~g_{ab,c}str
(D_\alpha \Phi ^a D^\alpha \Phi ^b \Phi ^c)+\hbox{fermion
terms},\ee with $g_{ab,c}$ the derivative of the classical bulk
metric.
 These terms lead to a coupling between the
worldvolume flavor current  $J_\alpha ^{[bc]}=tr (\Phi
^{[b}\partial _{\alpha} \Phi ^{c]})+$fermion terms (of course,
it's the fermion terms which are relevant for anomalies) and a
``bulk gauge field:'' \be \int d^{p+1}x J_\alpha ^{[bc]}A^\alpha
_{[bc]} \qquad \hbox{with}\qquad A^\alpha
_{[bc]}=g_{a[b,c]}\partial ^\alpha \phi ^a, \ee where $\Phi ^a$ is
a $U(K)$ adjoint and $\phi ^a$ is its eigenvalue, which we take to
be all the same for $K$ coincident branes.  Simply put, the
D-brane couples to angular momentum, which is gauged in the full
string theory.  The ``gauge field'' is the Christoffel symbol,
i.e. derivatives of the graviton, $g_{a[b,c]}$, pulled back to the
worldvolume via $\partial ^\alpha \phi ^a$.

The above coupling between the flavor current and the bulk closed
string modes goes away in the commutative $\alpha '\rightarrow 0$
decoupling limit.  However, in the NCFT noncommutative decoupling
limit, the worldvolume theory retains this coupling to the bulk
closed string modes in intermediate channels of non-planar loop
amplitudes. In particular, for our present case of interest, the
NCFT retains the information that the worldvolume flavor
symmetries correspond to bulk gauge symmetries.  Also retained is
the cancellation of all gauge anomalies, including that of the
gauged flavor symmetry.

Though our discussion can be generalized, we focus on a particular
example, non-commutative ${\cal N} =4$ $U(N)$ gauge theory. Both
the commutative and the non-commutative theory respect a $SU(4)_R$
flavor symmetry.  In the commutative case, the $Tr SU(4)_R^3$ 't
Hooft anomaly is non-zero: all of the gauginos are in the ${\bf
4}$, with none in the ${\bf \overline 4}$, so the 't Hooft anomaly
is $Tr SU(4)_R^3=|G|$, the dimension of the gauge group.  This is
an obstruction to gauging $SU(4)_R$ in the commutative theory; one
would first have to add some additional massless fields to cancel
this obstruction.  Though 't Hooft anomalies are generally
expected to remain unchanged by the dynamics of the theory, we
claim that making the theory noncommutative does effectively
change the 't Hooft anomaly, actually making it vanish.

In the full 10d string theory, before taking the decoupling limit,
the more precise statement about the cancellation of the 't Hooft
anomaly is that the $SU(4)_R$ current is indeed conserved, but
there is anomaly inflow of this current between the brane and the
bulk spacetime.  The natural holographic analog of this for the
NCFT is that current inflow into the bulk should correspond to
current flowing across energy scales.

In section 2 we compare commutative and non-commutative field
theory flavor anomaly calculations to string theory anomaly
calculations, and will demonstrate that the flavor anomalies
vanish in non-commutative field theories obtained from string
theory in the decoupling limit. In section 3, we argue that this
anomaly cancellation is in fact a consequence of the
reintroduction of closed string modes, which cause the symmetry to
be gauged.  This is outlined in the matrix model description of
NCFT.

\section{The Cancellation of Flavor 't Hooft Anomalies}

Though our discussion applies to any non-commutative gauge theory
which arises as a limit of string theory, we'll focus on the
particular case of non-commutative ${\cal N}=4$ $U(N)$ gauge
theory, which arises via D3-branes in the presence of a background
NS-NS B-field.  The non-commutative theory preserves the $SU(4)$
flavor symmetry.  Before taking the decoupling limit, the $SU(4)$
symmetry is a gauge symmetry, with the gauge field corresponding
to the graviton, as in eqn. (3).
(As seen in 10d IIB sugra on $S^5$
\cite{Kim:1985ez}, the $SU(4)_R$ gauge field actually comes {}from
a linear combination of the metric and the four-form gauge field.)

Consider the $Tr SU(4)_R^3$ 't Hooft anomaly first in the
commutative theory before taking the decoupling limit.  This
anomaly can be computed by a string theory world-sheet annulus
diagram which, as we'll discuss, yields the result that the
anomaly vanishes.  Of course, this is the expected result, given
that $SU(4)_R$ is a gauge symmetry before taking the decoupling
limit.  Now consider the commutative field theory in the
decoupling limit, which has a non-zero $Tr SU(4)_R^3$ 't Hooft
anomaly.  This non-zero result is obtained because, in the
commutative decoupling limit one omits the region of the moduli
space where the annulus becomes a long, thin, cylinder.  This
omitted part is the non-zero bulk inflow contribution to the
anomaly, coming from closed string modes in the bulk, which would
cancel the anomaly if included.

This is similar to the string theory cancellation of reducible
anomalies, which arise from non-planar diagrams.  These diagrams
are regularized by string theory, and thus there is no reducible
anomaly, even if the corresponding field theory does have one.
Again, the field theory limit omits the region of the annulus
moduli space where it's a long cylinder, which cancels the
anomaly. The space-time interpretation is that the long cylinder
is the closed string channel, which contains the tree-level
diagram involving the Green-Schwarz field responsible for
cancelling the anomaly.

Now NCFT, even in the decoupling limit, retains stringy behavior
in non-planar diagrams.  For example, the stringy automatic
Green-Schwarz anomaly cancellation of reducible anomalies occurs
in NCFT \cite{Intriligator:2001yu}.  We now argue that in NCFT the
$Tr SU(4)_R^3$ 't Hooft anomaly is also automatically cancelled,
even in the decoupling limit, by the ``bulk inflow'' closed string
contributions to the annulus diagram.   To do this, we consider
the full string theory anomaly diagram, and then take the
non-commutative field theory limit.  We will see that, unlike the
case of the commutative field theory limit, in the NCFT limit the
diagram is still finite and hence non-anomalous.

The $Tr SU(4)_R^3$ anomaly diagram is an annulus with three
insertions of the $SU(4)_R$ gauge field (essentially the bulk
graviton, as in eqn. (3)).  Because these gauge fields are closed
string modes, their vertex operators are to be inserted in the
bulk of the annulus diagram, at locations which are to be
integrated over. Fortunately, our main conclusion will not require
that we actually compute the diagram in detail.  Instead, we only
need to argue that it is regulated in the UV, as this is
sufficient to show that the $Tr SU(4)_R^3$ anomaly vanishes.  To
show that the diagram is regulated in the UV, we don't even need
to worry about specific form of the gauge boson vertex operator;
all that is needed is the form of the worldsheet propagator
between the closed string vertex operator insertions.

The worldsheet propagator is the Green function on the annulus,
with non-zero B-field and appropriate boundary conditions, which
was exhibited e.g.
in \cite{Liu:2000ad}, \cite{Chaudhuri:2000nz}.
The relevant term is
\bea {\cal G}^{\mu \nu}(\omega_1 , \omega_2 )&=& {(\Theta G
\Theta)^{\mu \nu} \over 8\pi^2 \alpha'} \left[\ln |{\nu_1
({\omega_1 - \omega_2 \over 2\pi \imath t} | {\imath \over t})
\over \nu_2 ({\omega_1 + \bar \omega_2 \over 2\pi \imath t} |
{\imath \over t}) }|^2 - {[Re \omega_1 - \pi]^2 + [Re \omega_2 -
\pi]^2 \over \pi t} +{2\pi \over t}\right] \nonumber\\ &+&
irrelevant\, terms, \eea where $t$ is the modulus of the annulus
and $\omega_{1,2}$ are the worldsheet bulk insertion points.  If
there is non-planar momentum $k_\mu$ flowing between these vertex
operator insertions, the annulus diagram will contain a factor of
$$\exp(-k_\mu k_\nu {\cal G}^{\mu \nu});$$
these closed string factors yield the stringy UV regulator for the
diagram, which imply that the diagram is finite, and thus anomaly
free, in the full string theory before taking the $\alpha
'\rightarrow 0$ decoupling limit.

We now consider the NCFT limit, $\alpha '\rightarrow 0$ with
$\alpha 't$ fixed.  The anomaly will still vanish because, even in
the decoupling limit, the above UV suppression factors continue to
provide a UV regulator for the anomaly diagram.  This is because
the above propagator, in the decoupling limit, continues to
contain terms of the form \be \propto {(\Theta ^2 )^{\mu \nu}
\over \alpha' t} \ee which lead to the UV regulator in the NCFT
limit.  There is indeed such a term in the above propagator
whenever the vertex operator insertion locations satisfy
$Re\,\omega _1\neq Re\,\omega _2$.  A special case of this is all
non-planar open string diagrams, where one vertex operator is on
one boundary of the annulus, $Re\, \omega _1 =0$, and the other is
on the other boundary of the annulus, $Re\, \omega _2=\pi$. But
the above UV suppression factor also occurs for closed string
vertex operators, inserted in the bulk of the annulus, as long as
$Re\, \omega _1\neq Re\, \omega _2$.

For our $SU(4)_R^3$ anomaly calculation, we have 3 closed string
vertex operators inserted at locations $\omega _i$, $i=1\dots 3$,
which should be integrated over the bulk of the annulus. As long
as the insertions are at different $Re \, \omega_i$, the diagram
will have the UV damping factors discussed above, and thus no
anomaly.  The integration over the $\omega _i$ will include
regions where the $Re\, \omega_i$ are not separated, and thus
without the UV damping, but such regions are a set of measure
zero.  Thus, the integrated diagram remains UV finite, and the
SU(4) R-symmetry anomaly is automatically canceled.  It is easy to
see that this generalizes to all global symmetries in
string-derived NCFT, since any global symmetry in such a theory
must arise from a gauge symmetry with a closed string gauge boson
in the underlying string theory.  The associated one-loop annulus
anomaly diagram with closed string insertions will be finite,
exactly as above.

Note that, although the usual $U(N)$ symmetry is non-commutative,
the $SU(4)_R$ gauge symmetry will be commutative. This is because
$SU(4)$ gauge boson is a closed string, so the argument in
\cite{Seiberg:1999vs} for noncommutivity, based on open string
gauge fields, does not apply.  Of course the $SU(4)_R$ symmetry
could not possibly have been noncommutative in any case, since
it's actually $Spin(6)$, with scalars in the ${\bf 6}$ and
fermions in the spinor ${\bf 4}$.

\section{Reappearance of the Bulk Graviton Couplings}

The cancellation of the flavor anomalies in the previous section
was via the closed string UV suppression factors of the form
$e^{{\alpha' \over t} k_{\mu} g^{\mu \nu} k_{\nu}}$, which do not
become trivial in the NCFT decoupling limit thanks to the scaling
of eqn. (1).  These suppression factors can be considered as a
stretched string effect \cite{Sheikh-Jabbari:1999vm} -
\cite{Liu:2000ad} as they can be written as $e^{\sim {\Delta x^2
\over \alpha' t}}$ \cite{Liu:2000qh}.

The anomaly cancellation is clearly very similar to that of the
full string theory before taking the decoupling limit.  The
natural interpretation, as in string theory, is that the annulus
diagram differs from the naive field theory diagram by including
the closed string propagation region of the moduli space, where
the cylinder has non-zero height.  Thus the anomaly is naturally
regarded as being cancelled by closed string bulk inflow,
involving closed string modes which are re-introduced by UV/IR
mixing.  This is analogous to the claim of
\cite{Intriligator:2001yu}, that the mixed anomalies are canceled
by the coupling of the non-commutative gauge bosons to closed
string axion-type fields, which cancel the anomaly by the
Green-Schwarz mechanism.  (A difference is that there the closed
string field is a twisted sector axion, living in the brane
worldvolume, whereas here the closed string modes generally live
in the bulk.)

So the $Tr SU(4)_R^3$ anomaly cancellation mechanism is naturally
the same as in the full string theory.  Moreover, the reason why
this actually had to be the case is also natrually the same as in
the full string theory: the flavor symmetries are secretly gauge
symmetries. This is because the closed string gauge bosons, which
couple to the flavor current as in eqn. (3), actually do not
decouple in the NCFT limit.  They do not exist as external,
on-shell states.  But, nevertheless, their presence is seen in
non-planar NCFT diagrams, which includes all closed string modes
in the intermediate channel via the propagators (or, in other
words, UV/IR mixing) discussed in the previous section. Because
NCFT retains the couplings, as in eqn. (3), to the closed string
gauge fields, the $SU(4)_R$ symmetry is secretly gauged.

This interpretation also explains why the observation of
\cite{Intriligator:2001yu}, that all ABJ type mixed flavor-gauge
anomalies automatically vanish (via Green-Schwarz cancellation) in
noncommutative field theories, also actually had to be the case.
If the flavor symmetries were really global symmetries, such
anomalies need not vanish.  But because the flavor symmetries are
secretly gauged, consistency of the theory requires these
anomalies, along with the 't Hooft anomalies, to all be cancelled.
String theory, and also the NCFT decoupling limit, automatically
ensure that this is the case.

The secret gauging of the $SU(4)_R$ symmetry, along with the
supersymmetry algebra, implies that gravity is also included and
one is discussing supergravity.  This is consistent with our
picture for NCFT, since all of the closed string modes, and in
particular the 10d supergravity fields, remain coupled as
off-shell, intermediate states in non-planar diagrams.  The fact
that 10d bulk closed string modes remain coupled via non-planar
diagrams also fits with the remark in \cite{VanRaamsdonk:2001jd},
that the leading non-planar one-loop effective action will lead to
a 10d, non-relativistic gravitational force between sources of
stress-energy.

The coupling of the worldvolume flavor current to the bulk can
also be seen in the matrix model description of NCFT.  As an
example, consider $Dp$ branes with $p$ even via the M-theory
matrix model (one can similarly consider $p$ odd in the IIB matrix
model).  Follow the discussion in \cite{Seiberg:2000zk}, we expand
the above action around a solution which captures the
noncommutivity: $X^I=X_0^I+Y^I$, with $X_0^i=x^i$ and
$X_0^{a+p}=0$, $Y^i= \theta ^{ij} \widehat A_j$, $Y^{a+p}=2\pi
\alpha ' \Phi ^a$, with $i=1\dots p$, $a=1\dots 9-p$, and
$[x^i,x^j]=i\theta ^{ij}$ (with $\theta ^{ij}$ of rank $p$).  For
flat $g_{IJ}$, this leads to the action for the NCFT Yang-Mills
theory in the worldvolume of the Dp brane \cite{Seiberg:2000zk}.

To see our couplings to the bulk gravitons, we should expand the
matrix model action for a general curved metric $g_{IJ}$
\cite{Douglas:1998zw}: \be\int dt\ STr\left({1\over
2}g_{IJ}(X)D_tX^ID_t
X^J-g_{IK}(X)g_{JL}(X)[X^I,X^J][X^K,X^L]+\dots \right),\ee where
$STr$ is the symmetrized trace, $\dots$ are the fermion terms, and
$I,J=1\dots 9$ run over the directions transverse to the D0s.
Expanding around the solution corresponding to NCFT, we replace
$$g_{IJ}(X)=\sum _{n=1}^\infty {1\over n!}g_{IJ, R_1\dots
R_n}(X_0) Y^{R_1}\dots Y^{R_n}$$ for all occurrences of the metric
in the above action.  Following \cite{Seiberg:2000zk}, this then
leads to the NCFT action, along with additional couplings, as in
\cite{Taylor:1999tv}, like
$$\int dt \sum _{n=0}^\infty {1\over n!}(T^{IJ(i_1\dots
i_n)}\partial _{i_1} \dots \partial _{i_n}g_{IJ}(0)+\dots),$$ with
$T^{IJ(i_1\dots i_n)}$ moments of the stress tensor. The $n=1$
term in the above yields our desired coupling eqn. (3) between the
worldvolume flavor currents and the associated bulk gauge field.
Even though there is no non-commutivity in the time direction,
$J_0^{[bc]}$ does couple to an associated bulk gauge field, as
seen upon expanding the kinetic term in the action (6).

Finally, we make  a general comment about the kinetic terms of the
closed string modes. In the NCFT limit, say for D3 brane with the
non-commutivity only in the $x^1$ and $x^2$ spatial directions,
$[x_1,x_2]=i\theta$, the closed string metric eqn. (1) will have
non-zero components only in the non-commutative directions.
Thus, the closed string modes with no momentum in the
non-commutative directions will decouple (these are a set of
measure zero).  Since the only closed string modes which couple
have momenta along the non-commutative directions, which are taken
to be only space directions, those modes will be space-like.  The
exponential factor $e^{{\alpha' \over t} k_{\mu} g^{\mu \nu}
k_{\nu}}$ appearing in the non-planar amplitude will thus have an
exponent which is always negative, and thus always damp the UV
limit of the integration over non-planar momenta.  Though there
are no finite energy poles associated with the closed string
degrees of freedom, the coupling in the noncommutative directions
between the closed string gauge boson modes and the flavor
currents are sufficient to ensure finiteness of the one-loop
diagram and invariance of the theory under the local flavor
symmetry transformations.

\vskip .2in {\bf Acknowledgments}

\vskip .1in We would like to thank J. Gomis, A. Rajaraman, L.
Susskind, M.M. Sheikh-Jabbari and M. Van Raamsdonk for useful
discussions and N.  Seiberg for several useful correspondences. J.
K. would like to thank the Aspen Center for Physics for its
hospitality.  This work is supported by DOE-FG03-97ER40546.

\end{document}